# Time-resolved observation of coherent excitonic nonlinear response with a table-top narrowband THz pulse wave

K. Uchida[1], H. Hirori[2,3,*], T. Aoki[3,4], C. Wolpert[2], T. Tamaya[2], and K. Tanaka[1,2,3]

[1] *Department of Physics, Graduate School of Science, Kyoto University, Kyoto 606-8502, Japan*

[2] *Institute for Integrated Cell-Material Sciences (WPI-iCeMS), Kyoto University, Kyoto 606-8501, Japan*

[3] *CREST, Japan Science and Technology Agency, Saitama 332-0012, Japan*

[4] *Department of Applied Physics, School of Advanced Science and Technology, Waseda University, Tokyo 169-8555, Japan*

T. Mochizuki, C. Kim, M. Yoshita, and H. Akiyama

*Institute for Solid State Physics, University of Tokyo, and JST-CREST, Chiba 277-8581, Japan*

L. N. Pfeiffer and K. W. West

*Department of Electrical Engineering, Princeton University, New Jersey 08544, USA*

By combining a tilted-pulse-intensity-front scheme using a LiNbO$_3$ crystal and a chirped-pulse-beating method, we generated a narrowband intense terahertz (THz) pulse, which had a maximum electric field of more than 10 kV/cm at around 2 THz, a bandwidth of ~50 GHz, and frequency tunability from 0.5 to 2 THz. By performing THz pump and near-infrared probe experiments on GaAs quantum wells, we observed that the resonant excitation of the intraexcitonic 1s-2p transition induces a clear and large Autler-Townes splitting. Our time-resolved measurements show that the splitting energy observed in the rising edge region of electric field is larger than in the constant region. This result implies that the splitting energy depends on the time-averaged THz field over the excitonic dephasing time rather than that at the instant of the exciton creation by a probe pulse.

*hirori@icems.kyoto-u.ac.jp

Excitons in semiconductors, which are hydrogen-like quasi-particles energetically scaled by a factor of 1/1000 with respect to atomic systems, are fascinating systems for studying light–matter interactions in the presence of Coulomb interactions.[1] The intraexcitonic transition energy lies in the THz frequency range and has a large dipole moment.[2,3] Narrowing and shaping techniques of THz pulse can be utilized for coherent manipulation of excitonic quantum systems,[4-6] and time-resolved experiments with THz pulse systems generating phase-stable transients may provide new insights into intriguing nonlinear phenomena, e.g., excitonic sideband generations and high-order harmonics.[7,8] Moreover, these studies may indicate the potential applications of THz-induced nonlinear optical effects for the purpose of quantum information and ultrahigh-speed optical signal processing.[9] So far, several techniques using table-top Ti:sapphire-based lasers to generate narrowband intense THz waves have been developed,[10] e.g., optical rectification in periodically poled lithium niobate (PPLN) and optical rectification of spatially or temporally shaped ultrafast optical pulse in a bulk LiNbO$_3$ crystal.[11-15] However, the THz wave sources with intense THz field strengths (>10kV/cm) have broad spectral bandwidths (>0.1 THz), thus obscuring the observation of excitonic coherent nonlinear interactions.

In this study, we temporally modulated the optical pulse intensity at THz frequencies by using a chirped-pulse-beating method,[6,16,17] and combining it with a tilted-pulse-intensity-front scheme,[18,19] generated intense narrowband THz pulses by optical rectification of quasi-sinusoidal optical pulses in LiNbO$_3$ crystal. By performing the THz pump and near-infrared (NIR) probe spectroscopy, we observed that the incident THz pulse induced a strong spectral modulation of the 1s heavy-hole exciton peak of the GaAs quantum wells through Autler-Townes (AT) splitting.[9,14,20,21] Our time-resolved measurements show that the splitting energy observed in the rising edge region of electric field is larger than in the constant region. This result implies that the splitting energy depends on the time-averaged THz field over the excitonic dephasing time rather than that at the instant of the exciton creation by a probe pulse.

Figure 1 schematically illustrates the experimental setup for the tunable narrowband THz generation, and the THz pump and NIR probe experiment. The laser source for the THz generation was a Ti: sapphire regenerative amplifier (repetition rate of 1 kHz, center wavelength of 800 nm, pulse duration of 100 fs (FWHM), and pulse energy of 4 mJ/pulse). Using the tilted-pulse-intensity-front scheme,[18,19] we generated intense narrowband THz pulses by optical rectification of quasi-sinusoidal optical radiation modulated at THz frequencies in LiNbO$_3$. To enhance the THz generation efficiency, the crystal was cooled down to 80 K.[22] The experimental setup for producing optical modulation was similar to that in other studies and is schematically illustrated in Fig. 1.[6,16,17] Such pulses can be prepared when optical pulses, whose linear chirp is controlled by a grating pair, pass through a Michelson interferometer. The optical rectification process in LiNbO$_3$ crystal converts the interferometer-output into THz radiation. The center frequency of the THz pulse $\omega_{THz}$ is given by

$$\frac{\omega_{THz}}{2\pi} = -\frac{\tau}{4\pi\Phi^{(2)}}, \quad (1)$$

$$\Phi^{(2)} = -\frac{4\pi^2 cb}{\omega_0^3 d^2}\left\{1 - \left[(\frac{2\pi c}{\omega_0 d}) - \sin\gamma\right]^2\right\}^{-3/2}, \quad (2)$$

where $\tau$ is the time delay at the Michelson interferometer, and $\Phi^{(2)}$ is the group delay dispersion at the grating pair around the center frequency of the input NIR pulse $\omega_0$. Here, $c$ is the speed of light, $\gamma$ is the incident angle at the grating, $b$ is the perpendicular distance between gratings, and $d^{-1}$ is the grating groove density. The center frequency of the resultant THz pulses change with these parameters, and in this study, we set them to be $b$=5 cm, $\gamma$=39º, and $d^{-1}$=1800 mm$^{-1}$.

Figure 2(a) shows the temporal profile of a THz pulse centered around 2 THz, as measured by electro-optic sampling (EOS). The maximum electric field amplitude reached 10 kV/cm in air. Figure 2(b) shows the normalized power spectra of THz pulses with different center frequencies (typical spectral bandwidth of ~50 GHz) along with their field strengths. The frequency could be changed according to the relation of Eq. (1), and this frequency tunability was confirmed between 0.5 THz to 2 THz, as shown in Fig. 2(c). The $\Phi^{(2)}$ (=−7.2 ×10$^{-25}$ s$^2$) deduced from the fitting of the experimental results with Eq. (1) (Fig. 2(c)) was in good agreement with the value ($\Phi^{(2)}$ =−7.3×10$^{-25}$ s$^2$) calculated from Eq. (2).

By using the generated THz pulse as a pump pulse, we performed THz-pump and NIR-probe experiments on GaAs quantum wells (QW). The QW sample consisted of a stack of ten 12-nm-wide undoped GaAs wells separated by 10-nm-wide Al$_{0.3}$Ga$_{0.7}$As barriers, and it was glued to a SiO$_2$ substrate for the optical transmission experiments. The sample was mounted in a vacuum on a liquid-helium-cooled cold finger and was cooled down to 10 K. Our QW were designed so that the NIR spectrum would cover a range around the 1s heavy-hole (HH) exciton peak of the sample. The NIR-pulse for the EOS was also used for the probe pulse (exciton density $N_{ex}$ generated by the NIR-probe ~10$^{10}$ cm$^{-2}$). This enabled us to examine the relation between the pump THz electric field detected by EOS and the time evolution of the absorption change measured by the pump-probe scheme with a time resolution of 100 fs. After passing through the sample, the NIR-probe-pulse was resolved by a spectrometer (energy resolution ~ 0.3 meV) and detected by a charge-coupled-device camera.[23] The spots of the THz-pump and NIR-probe pulses at the sample position were respectively ~200 and ~50 μm in diameter.

Figure 3(a) shows the probing scheme of AT (or Rabi) splitting absorption in our THz-pump and NIR-probe experiment. When the NIR pulse alone excites the QW (left side), one may generate a 1s-exciton polarization which coherently oscillates with the 1s polarization frequency and decays with the dephasing time $T_2$, and its power spectrum corresponds to the absorption resonance. When the intra-excitonic 1s-2p transition in the QW is simultaneously excited by a THz pulse (right side), the additional coupling alternates the polarizations between the 1s and 2p-exciton at the Rabi frequency ($\Omega_R$). In this case, the power spectrum shows AT splitting that can be detected as new resonances in the NIR absorption spectra.[20,21]

Figure 3(b) shows part of the temporal and spectral profiles of the pump THz pulse at the sample position. Here, we defined the electric field amplitude $E_{THz}(t)$ as the envelope, i.e., $E(t)=E_{THz}(t)\cos(\omega_{THz}t)$, where $E(t)$ is the oscillating electric field. The photon energy (frequency) of the pump THz pulse $\hbar\omega_{THz}$ was set at 8.6 meV ($\omega_{THz}/2\pi=2.1$ THz) to nearly resonantly excite the intraexcitonic 1s-2p transition ($\hbar\omega_{12}=8.2$ meV) of the sample. Figure 3(c) shows the absorption spectra of the sample modulated by the pump pulse as a function of the time delay $T$ between the pump and probe pulses. Figure 3(d) shows the absorption spectra sliced at fixed times $T=-4.4$ ps (3.2 kV/cm) and $-0.4$ ps (5.2 kV/cm) in Fig. 3(c), and no THz excitation. The optically bright 2s state lies very close to the 2p state.[25] As shown in Figs. 3(b)-(d), the excitonic absorption peak splits in two and the splitting energy becomes large with an increase of electric field.[26] The modulated absorption can be expressed by a semi-empirical equation (Eq. (3)) that assumes the absorption spectra consist of lower and upper branches of the split 1s-HH exciton. The third term approximately includes the contribution from the higher excitonic states (2s, 3s, 4s,…) as well as the band-to-band transition for a two-dimensional (2D) density of states:

$$\alpha(\omega) = \frac{\alpha_{LB}\Gamma_{LB}^2}{(\omega-\Omega_{LB})^2 + \Gamma_{LB}^2} + \frac{\alpha_{UB}\Gamma_{UB}^2}{(\omega-\Omega_{UB})^2 + \Gamma_{UB}^2} + \frac{\alpha_c}{1+\exp\left(\frac{\Omega_c - \omega}{\Gamma_c}\right)}, \qquad (3)$$

where $\alpha_i$, $\hbar\Omega_i$, and $\hbar\Gamma_i$ (i = LB for the lower branch, UB for the upper branch, and c for the continuum) are the amplitudes of the corresponding functions, absorption-peak energies, and absorption widths, respectively. By fitting the observed spectra to Eq. (3) (see Fig. 3(d)), we obtain a splitting energy given by $\varepsilon_S = \hbar\Omega_{UB} - \hbar\Omega_{LB}$.

Figure 4(a) shows the field dependence of $\varepsilon_S$ obtained from the measured absorption spectra where the electric field amplitude was almost constant ($T=-0.4$ ps) for the THz pulses with different peak amplitudes (a pair of wiregrid polarizers in Fig. 1 was used to change the amplitude). The splitting energy $\varepsilon_S$ is proportional to $E_{THz}$, and this linear relationship indicates that the splitting of the 1s excitonic peak originates from AT splitting, since the AT splitting energy $\hbar\Omega_R$ is proportional to $E_{THz}$, i.e., $\hbar\Omega_R = d_{12}E_{THz}$. Here, $d_{12}$ is the dipole moment of the 1s-2p transition, and was deduced from the experiment ($=\hbar\Omega_R/E_{THz}=112$ eÅ).[27] The maximum splitting energy of ~7 meV is comparable to $\hbar\omega_{12}$ (=8.2 meV), indicating that the electric field of our THz pulse source leads to a highly nonlinear interaction ($\hbar\Omega_R \approx \hbar\omega_{12}$).

Our time-resolved experiment allowed us to obtain the excitonic dephasing time $T_2$ relevant to the dynamics of coherent interactions. To extract $T_2$ under THz irradiation, as shown in Fig. 4(b), we compared the splitting energies $\varepsilon_S$ in the two time-delay regions, one where the electric field amplitude rises transiently ($|T| \gtrsim 2$ ps) and one where it is almost constant and regarded as CW-like ($|T| \lesssim 2$ ps) (Fig. 4(c)). As shown in Fig.

4(b), the $\varepsilon_S$ (=2.1 meV) in the CW-like case ($T$=−0.4 ps, red line) was smaller than that (=2.7 meV) in the transient case ($T$=−7.2 ps, blue line) despite having the same $E_{THz}$ (=1.9 kV/cm), and rather was almost the same with that for the smaller $E_{THz}$ (=1.3 kV/cm, green line) at $T$=−8.3 ps.[28]

The THz-perturbed excitonic polarization appeared as AT splitting in the absorption spectra, and the perturbation continued during the $T_2$ of the polarization.[29,30] Thus, we assumed that $\varepsilon_S$, instead of the instantaneous electric field amplitude $E_{THz}$, depends on the following time-averaged electric amplitude $E_{Avg}$, i.e., $\varepsilon_S = d_{12} E_{Avg}$:

$$E_{Avg}(T) = \int_T^\infty \frac{1}{T_2} \exp(-\frac{t'-T}{T_2}) E_{THz}(t') dt' . \qquad (4)$$

Note that $E_{Avg}$ is the time-averaged value of $E_{THz}(t)$ weighted with the exponentially decaying polarization. Thus, $E_{Avg}$ for the transient case becomes larger than the $E_{THz}$, whereas that for the CW-like case almost coincides with the $E_{THz}$. By using the $E_{Avg}$ given by Eq. (4), we may compensate for the displacement of the electric fields for the transient case and the CW-like one (blue and green lines in Fig. 4(b)), and derive $T_2$. Here, the $T$=−8.3 ps and $E_{Avg}$=1.9 kV/cm were used in Eq. (4), and $T_2$ was treated as a fitting parameter. The derived $T_2$ of 0.6 ps (±0.2 ps) at 1.9 kV/cm with Eq. (4) is similar to those derived from the spectral widths ($1/\Gamma_{LB}$ ~0.8 ps and $1/\Gamma_{UB}$ ~0.5 ps), showing that the THz generation system enables us to capture the sub-picosecond dynamics of excitonic coherent interactions. Figure 4(e) shows the temporal profiles of $E_{THz}$ and $\varepsilon_S$. As indicated by the arrows in Fig. 4(e), for the same $E_{THz}$ (=2 kV/cm) the $\varepsilon_S$ in the rising edge is larger than that in the falling edge, also implying that $\varepsilon_S$ depends on the $E_{Avg}$ rather than $E_{THz}$ at the instant of the exciton creation by a probe pulse.

In conclusion, on the basis of a tilted-pulse-intensity-front scheme and a chirped-pulse-beating method, we generated a narrowband intense THz pulse. By using the THz pulse as a pump pulse, we observed that the resonant excitation of the intraexcitonic 1s-2p transition in GaAs QW induces a clear and large AT splitting. Our time-resolved measurements allowed us to capture the sub-picosecond dynamics of excitonic coherent interactions, indicating that $\varepsilon_S$ depends on the time-averaged THz field over the excitonic dephasing time rather than that at the instant of the exciton creation. The narrowband intense THz pulses developed here have the possibility of being applied to studies of other quantum systems such as subband transitions in nanostructured semiconductors and impurity Rydberg states in ubiquitous semiconductors.[9,31,32]

**Acknowledgements**

We are grateful to Ryo Shimano for fruitful discussions. This study was supported by KAKENHI (26286061) from JSPS, Industry-Academia Collaborative R&D and from Japan Science and Technology Agency (JST). Part of this work was supported by the KAKENHI (26247052) from JSPS and JST-PRESTO. The work at Princeton University was funded by the Gordon and Betty Moore Foundation through the EPiQS initiative Grant GBMF4420, and by the National Science Foundation MRSEC Grant DMR-1420541.

solid line), maybe because of the influence of the higher energy levels (Ref. [21]).

**Figure caption**

**Fig. 1** Schematic setup of narrowband THz generation, and THz-pump and NIR-probe setup. The probe pulse was directed through a small hole in the parabolic mirror used to focus the THz pulse. BS: non-polarized beam splitter, HWP: half wave plate, W1: glass (BK7) window, W2: plastic (Tsurupica) window. EOC indicates a GaP or ZnTe crystal used in the electro-optic sampling (EOS).

**Fig. 2** (a) The expanded THz pulse in air between −5 and 5 ps measured by the EOS (center frequency is around 1.9 THz) with a 1-mm-thick ZnTe crystal. The inset shows its overview between −15 and 15 ps. (b) Normalized power spectra of THz pulses with different center frequencies. Red open circles show the maximum electric fields of their THz pulses. (c) Red circles show the experimental results of the THz center frequency dependence on the time delay $\tau$ shown in Fig. 1. The dashed line is the calculation with Eqs. (1) and (2). The uncertainties of center frequency and electric field are typically ±1% and ±5%.

**Fig. 3** (a) Schematic energy diagram of the intraexcitonic transitions in the THz-pump and NIR-probe experiment. (b) The temporal profiles of electric field and amplitude of THz pulse in the sample measured by the EOS with a 400-μm-thick GaP crystal, and calibrated by taking into account the THz multiple reflection.[24] The inset shows its power spectrum. (c) Absorption (−log(Trans.)) spectra under THz irradiation corresponding to Fig. 3(b) as a function of time delay $T$ between pump and probe pulses. $T$=0 indicates the peak position of the THz-pump pulse in the time domain. Trans. indicates the transmittance of the sample in which the background term due to optical multiple reflection interference in the sample has been removed. (d) Measured absorption spectra with and without THz pump (the gray filled areas and black solid line). The red, blue, and black dashed lines respectively show the components of UB, LB, and total of the fit with Eq. (3). The spectra are vertically offset for clarity.

**Fig. 4** (a) The electric field $E_{THz}$ in the sample dependence of the energy splitting $\varepsilon_S$ (solid circle) and linear fit with $\hbar\Omega_R=d_{12}E_{THz}$ (dashed line). The uncertainties of $\varepsilon_S$ in the fitting are typically ±10%. (b) Absorption spectra for the internal electric field amplitude (1.9 kV/cm and 1.3 kV/cm) at the different time delays $T$ (=−7.2, −0.4, and −8.3 ps) as indicated by the arrows in (c). Dashed lines show fits with Eq. (3). The dotted lines in (b) and (c) are guides for eye. (c) Temporal profiles of THz amplitude with two different peak fields (1.9 and 6.8 kV/cm). (d) Schematic diagram of THz pulse and 1s-excitonic polarization $P_{1s}$ induced by the NIR-probe pulse. $E_{THz}$ and $E_{Avg}$ respectively indicate the electric field amplitudes of the THz pulse and the time-averaged one derived by Eq. (4). (e) Temporal profiles of the $\varepsilon_S$ and $E_{THz}$ which are shown from their half maximum to maximum values. The grey filled area corresponds to the $E_{THz}$ (right axis), and the blue dashed line to the $\varepsilon_S$ (left axis). The two vertical arrows show the $\varepsilon_S$ in the rising and falling edges at the same $E_{THz}$ (=2 kV/cm).

**Fig.1**

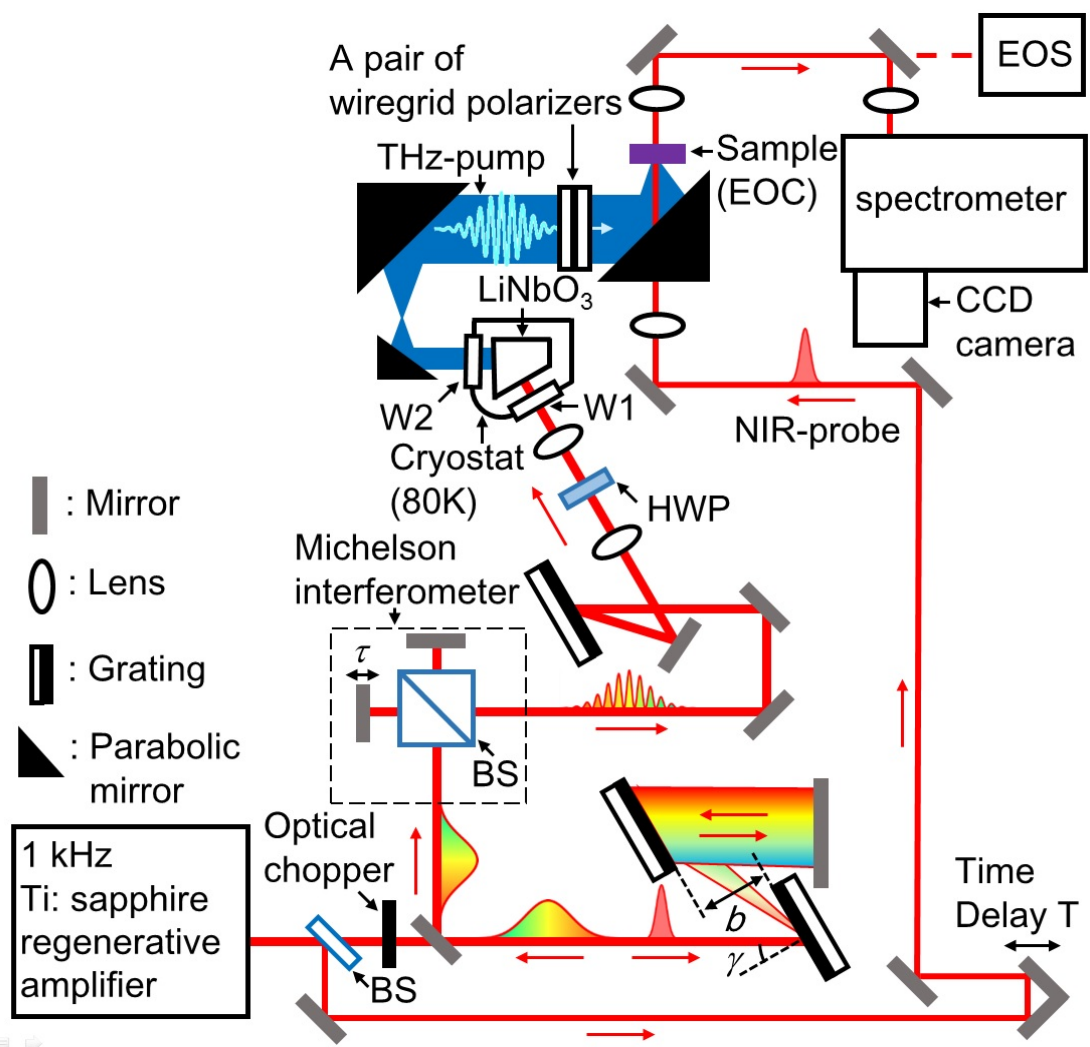

K. Uchida

**Fig.2**

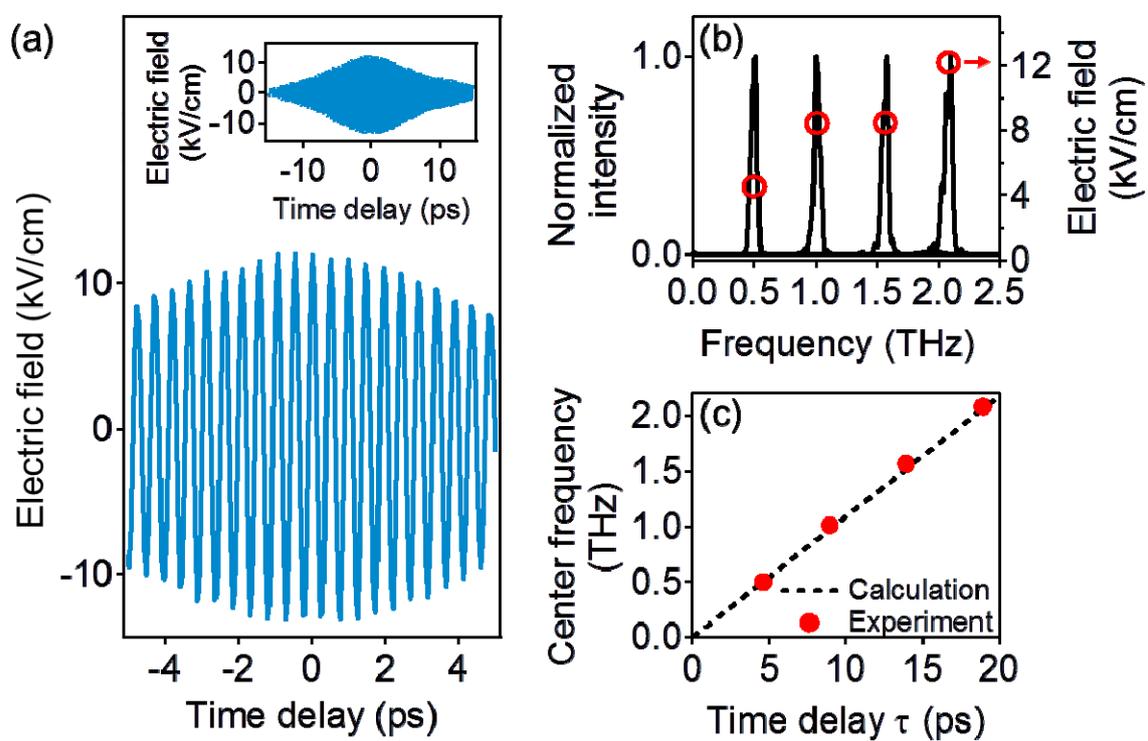

K. Uchida

**Fig. 3**

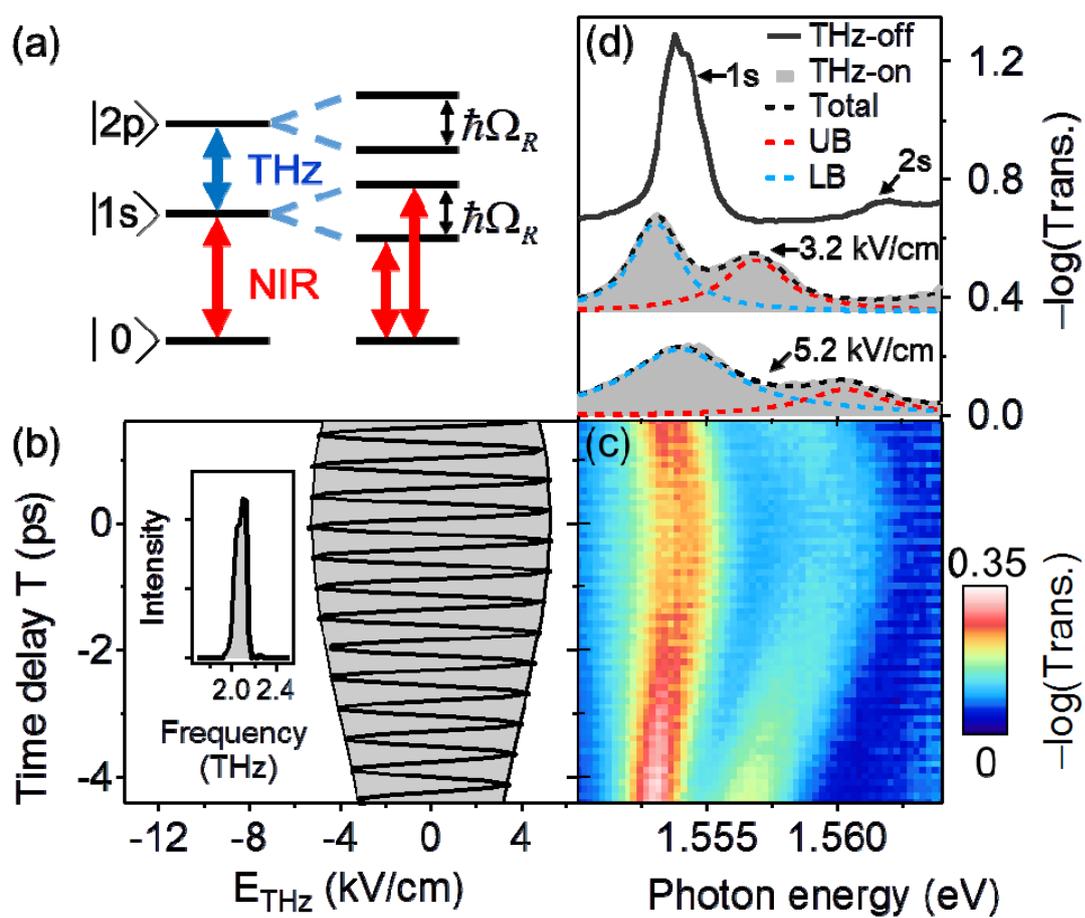

K. Uchida

Fig. 4

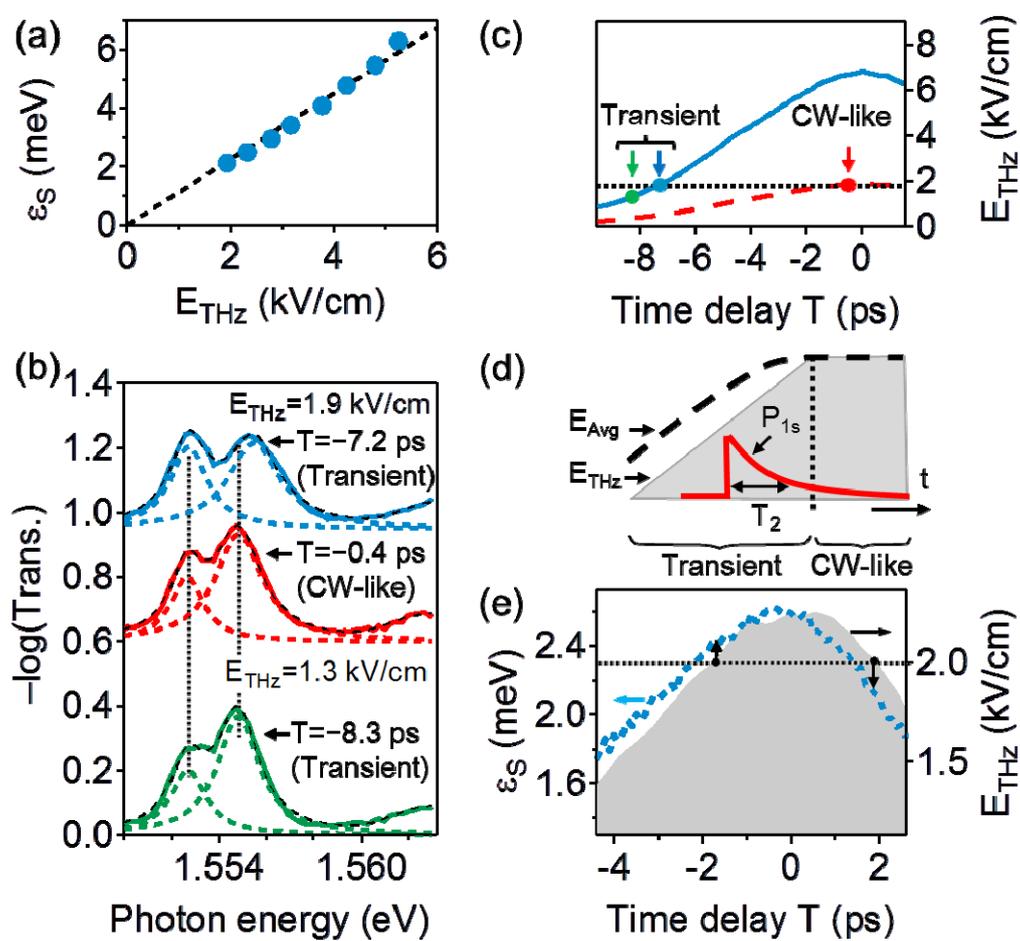

K. Uchida